\documentclass[fleqn,oneside]{article}
\usepackage{sectsty}
\usepackage{mathrsfs,amsmath,amssymb}%,times}

\newcommand{\cc}{\mathrm{c.c.}}
\newcommand{\rmi}{\mathrm{i}}
\newcommand{\rmd}{\mathrm{d}}
\newcommand{\Or}{\mathrm{O}}
\newcommand{\bea}{\begin{eqnarray}}
\newcommand{\eea}{\end{eqnarray}}
\newcommand{\beag}{\begin{eqnarray*}}
\newcommand{\eeag}{\end{eqnarray*}}
\newcommand{\be}{\begin{equation}}
\newcommand{\ee}{\end{equation}}
\newcommand{\beg}{\begin{equation*}}
\newcommand{\eeg}{\end{equation*}}
\newcommand{\iz}{\,{\scriptstyle{\wedge}}\,}
\newcommand{\frec}[2]{{\textstyle{\frac{#1}{#2}}}}

\newcommand{\heq}{\,\,\hat{=}\,\,}

\newcommand{\const}{\mathrm{const}}

\newcommand{\bxi}{\bar{\xi}}
\newcommand{\bL}{\bar{L}}
\newcommand{\bW}{\bar{W}}
\newcommand{\bY}{\bar{Y}}
\newcommand{\bpartial}{\bar{\partial}}
\newcommand{\hP}{\hat{P}}
\newcommand{\hu}{\hat{u}}
\newcommand{\cov}{{\scriptscriptstyle |}}
\newcommand{\zwe}{-\!\!\!\lower-.240em\hbox{$\lrcorner$}\,}
\newcommand{\scri}{\mathscr{I}^+}
\allsectionsfont{\normalsize}
\begin{document}
\title{{Kerr--Schild metrics and\\ gravitational radiation}}
\author{{\normalsize W{\l}odzimierz Natorf$^1$}}
\date{}
\maketitle
\begin{center}\textit{
$^1$Institute of Theoretical Physics,\\
Warsaw University,\\
Ho\.za 69, 00--681 Warsaw, Poland.\\}
\end{center}
e-mail: \texttt{nat@fuw.edu.pl}
\begin{abstract}\noindent We
describe conditions assuring that the Kerr--Schild type solutions
of Einstein equations with pure radiation fields are asymptotically
flat at future null infinity. Such metrics cannot describe ``true''
gravitational radiation from bounded sources---it is shown that the
Bondi news function vanishes identically. We obtain formulae for the
total energy and angular momentum at $\scri$. As an example we
consider non-stationary generalization of the Kerr metric given by
Vaidya and Patel.  Angular momentum and total energy are expressed
in closed form as functions of retarded time.\end{abstract}
\noindent {\sf {\small keywords: Kerr--Schild metrics, gravitational
radiation, Bondi--Sachs metrics.}}
\section{Introduction}
Metrics of the form\be\label{kerr-schild}\tilde{g}_{\alpha\beta}=
\eta_{\alpha\beta} + 2h\,\kappa_{\alpha}\kappa_{\beta},\ee where $\eta$
is a Minkowski metric, $\kappa$ is a null form and $h$ is a function,
were first studied by Trautman \cite{TR}. The main motivation for
this research was to describe the propagation of information by
gravitational waves. Kerr and Schild \cite{KS} reduced the Einstein
field equations and gave some explicit solutions, in particular the
well known Kerr metric. Later, G{\"u}rses and G{\"u}rsey \cite{GG}
have simplified the field equations to a single linear equation.
Recently, Kerr--Schild Ansatz and its modifications has been
examined in \cite{CHS,HI}, in particular, a systematic study in
an arbitrary number of dimensions has been performed. Metrics
\eqref{kerr-schild} satisfying the Einstein equations with a
cosmological constant have been examined in \cite{GS}.
The Kerr--Schild fields turned out to be useful also in more exotic
applications, such as to describe an accelerating dyon \cite{CA}.

\medskip

In this paper our interest is focused on finding physical properties
of the Kerr--Schild solutions of Einstein equations with pure
radiation fields,
\be\label{eqpr} \tilde{R}_{\alpha\beta}=
\Phi\kappa_{\alpha}\kappa_{\beta}.\ee
In this case the metric $\tilde{g}$ is algebraically special,
$k=\tilde{g}^{-1}(\kappa)$ being the degenerate eigenvector of the
Weyl tensor \cite{KS}.  In order to obtain formulae for the total
energy and angular momentum, we assume that the metric is
asymptotically flat at $\scri$ in sense of \cite{TB} and we find its
approximate Bondi--Sachs form. It turns out that none of the
Kerr--Schild metrics describes ``true'' gravitational radiation---the
Bondi news function always vanishes. The energy or angular momentum
loss can be nonzero only due to nonzero value of $\Phi$ in
\eqref{eqpr}. This result generalizes that of Cornish and Micklewright
\cite{CM} who examined the energy--momentum pseudotensor of the
twist-free ``photon rocket'' solutions \cite{KIN} and proved that the
source energy loss can be nonzero only due to the presence of the null
field radiated by the ``rocket''.  Kramer and von der G\"o{}nna
\cite{KGpr} proved the same property of Kinnersley's metrics using
approximate Bondi--Sachs coordinates. Their result states that the
``photon rockets'' are the only axisymmetric Robinson--Trautman
metrics without gravitational radiation.

As an example we consider ``radiating Kerr'' metric of Vaidya and
Patel \cite{VP}. Kramer and von der G\"o{}nna \cite{KG} have
considered the same metric using other coordinate system (the
``radiating Kerr'' solution has been published also by Kramer in 1972
\cite{KRRK}, few months before the publication of \cite{VP}.) In our
considerations we use the coordinates from \cite{VP} which happen to
coincide with some Bondi coordinates on scri. Our approach to angular
momentum seems to be more general than that in \cite{KG} where the
authors made use of the axial symmetry of the metric, obtaining
$\vec{L}$ out of the Komar integral.  In contrary to \cite{KG} we give
the explicit formula for energy.

\medskip

We use the same definition of asymptotic flatness at $\scri$ as in
\cite{TB} and \cite{TN}. The notation aims to be as close as possible
to that of \cite{TN}. The Penrose conformal factor is denoted by
$\Omega$, an equality on $\scri$ is denoted by $\heq$. Indices $A,B$
of tensors on the sphere $S_2$ are raised or lowered by means of minus
the standard metric $s_{AB}$ on $S_2$.

\medskip

If a spacetime with metric $\tilde g$ is asymptotically flat at
$\scri$ there exists a set of preferred coordinates $(u',r',x'^A)
_{A=2,3}$ (the Bondi--Sachs coordinates) such that \cite{TW}
\be\tilde g=\rmd u'(\tilde g_{00}\rmd u'+2\tilde g_{01}\rmd r'+
2\tilde g_{0A} \rmd x'^A)+ r'^2g_{AB}\rmd x'^A\rmd x'^B.
\label{BSmetric}\ee
The metric components have the following expansions in the negative
powers of $r'$:
\be\label{BS}\begin{split}\tilde g_{00}&=1-2M'r'^{-1}+\Or (r'^{-2}),\\
\tilde g_{01}&=1+ \Or (r'^{-1}),\\
\tilde g_{0A}&=\psi_A+\kappa_A r'^{-1} + \Or (r'^{-2}),\\
g_{AB}&=-s_{AB}+n_{AB}r'^{-1}+\Or (r'^{-2}),\end{split}\ee
where \be\psi_A=-\frac{1}{2}n^B_{\,\ A|B}\ ,\quad\det
g_{AB}=\det s_{AB}\label{BSeq} \ee
Coordinates $u'$ and
$r'$ are interpreted as a retarded time and the luminosity distance,
respectively. The function $M'$ is referred to as the Bondi mass
aspect. The $u$--derivative of the tensor $n_{AB}$ corresponds to the
Bondi news function and the form $\kappa_A$ (the angular momentum
aspect) is used in all definitions of angular momentum at future null
infinity.

Total energy--momentum vector and total angular momentum tensor can be
expressed as integrals over $S_2$ of expressions built out of $M'$,
$\psi_A$, $n_{AB}$ and $\kappa_A$. To define the energy it is
sufficient to know the modified Bondi mass aspect $\hat M$ defined by
\be\label{mhatdef}\hat{M}=M'+\frac{1}{4}n^{AB}_{\,\,\,\,\,\,\,\,|AB}.\ee
(see e.g. \cite{TB,JEZ}.) The Bondi energy--momentum corresponding to a
$u'=\const$ section of $\scri$ is given by
\be\label{BondiP}P^{\alpha}(u')=\frac{1}{4\pi}
\int\limits_{S_2}^{}\hat{M}l^{\alpha}\rmd\sigma,\ee
where $l=(1,\hat r)$ and $\hat r$ represents a point of $S_2$ in $R^3$. 

Assume that the following approximate Einstein equations hold on scri:
\be\label{amric}\Omega^{-1}\tilde{R}_{\alpha\beta}\heq
q\Omega_{\cov\alpha}\Omega_{\cov\beta},\ee
where $q$ is a function independent of $\Omega$.
One of the field equations \eqref{amric} at $\scri$ reads
\be\dot {\hat M}\heq -\frac{1}{8}\dot n^{AB}\dot n_{AB}
- \frac{1}{2}\Omega^{-2}\tilde R_{00}.\label{mloss} \ee
Integrating both sides of \eqref{mloss} over $S_2$ one obtains Bondi's
energy loss formula with a linear combination of terms describing
the loss of energy caused by gravitational radiation and radiation
of matter, respectively.

Definition of total angular momentum tensor at instant $u'$ \cite{TN} reads
\be L^{\alpha\beta}(u')= -\frac{3}{8\pi} \int\limits_{S_2}^{}
l^{[\alpha}_{\phantom{A}}l^{\beta]}_{\ \,\,,A}\hat\kappa^A \rmd\sigma\
\label{angm},\ee
where \be\label{kappahat}\begin{split}
&\hat\kappa_A=\kappa_A -\frac{1}{16}n^2,_A+
p(\frac{1}{24}n^2,_A-\frac{1}{3}n_{AB}\psi^B),\\
& n^2 = n_{AB}n^{AB},\ p=\const,\end{split}
\ee
(preferably $p=1$.) This formula corresponds to standard spinor
definitions of angular momentum at $\scri$ (see \cite{DS} and
references therein.) The procedure of finding
$\hat{M}$ and $\kappa_A$ is described in detail in \cite{TB} and \cite{TN}.

\section{Basic properties of the Kerr--Schild metrics}
The Kerr--Schild metrics are closely related to shear-free null
geodesic congruences in Minkowski spacetime \cite{KS}.  One can
rewrite the Minkowski metric $\eta=2\rmd u'\rmd v'-2\rmd\xi'
\rmd\bxi'$ as
\be\label{mink1}\eta = 2\kappa\rmd v'-2(\rmd\xi'+Y\rmd v')
(\rmd\bxi'+\bY\rmd v'),\ee
where
\be\label{kappa}\kappa=\rmd u'+Y\bY\rmd v' + \bY\rmd\xi'+Y\rmd\bxi'\ee
and $Y$ is a complex function of four variables. The Kerr theorem
states that in the analytic case the vector field corresponding to
$\kappa$ is tangent to a shear-free null geodesic congruence if and
only if $Y$ satisfies the equation
\be\label{Kerr}F(Y,u'+Y\bxi',\xi'+Yv')=0,\ee
$F$ being some holomorphic function of three variables.
Introducing new coordinates
\be\label{kscoord}\begin{split}
u&=u'+Y\bY v'+ Y\bxi' +\bY\xi',\\ \xi&=Y,\\ v&=v'\end{split}\ee
and defining
\be\label{defL}L(u,\xi,\bxi)=-\bxi'-v\bxi\ee
one obtains
\be\label{mksmetr}
\eta=2\kappa\omega-2(v+\partial\bL)(v+\bpartial L)\rmd\xi\rmd\bxi,\ee
where 
\be\label{kong}
\kappa=\rmd u + L\rmd\xi +\bL\rmd\bxi,\ee
\be
\omega=-L,_u\bL,_u\kappa + \rmd v -(v+\partial\bL)L,_u\rmd\xi-
(v+\bpartial L)\bL,_u\rmd\bxi,\ee
\be\label{de}
\partial=\partial_{\xi}-L\partial_u .\ee
Adding the term $2h\kappa^2$ to the flat metric
leads to the Kerr--Schild metric,
\be\label{kerrschild}\begin{split}
\tilde{g}&=2\kappa\left[
(h-L,_u\bL,_u)\kappa+\rmd v -
(v+\partial\bL)L,_u\rmd\xi-(v+\bpartial L)\bL,_u\rmd\bxi\,\right]-\\
&-2(v+\partial\bL)(v+\bpartial L)\rmd\xi\rmd\bxi.\end{split}\ee
Note that from \eqref{Kerr} and \eqref{defL} it follows that \be\label{dl=0}
\partial L=0.\ee

\section{Asymptotic flatness of the Kerr--Schild metrics}
In \cite{TN} we have formulated conditions assuring that
the twisting algebraically special metric satisfying
\eqref{eqpr} is asymptotically flat at $\scri$. 
These metrics are expressed in terms of the coordinates $(u,r,\xi,\bxi)$:
\be\tilde{g}=2\kappa(H\kappa+\rmd r+W\rmd\xi+\bW \rmd\bxi)-
2\frac{r^2+\Sigma^2}{P^2}\rmd\xi\rmd\bxi,\label{als_m}\ee
where
\be\label{als_c}\begin{split}
&\kappa=\rmd u+L\rmd\xi+\bL\rmd\bar\xi,\\
&W=-(r+\rmi\Sigma)L,_u+\rmi\partial\Sigma,\\
&H=-r(\log P),_u-\frac{mr+M\Sigma}{r^2+\Sigma^2}+         
P^2\mathrm{Re}[\partial(\bpartial\log P-\bL,_u)],\\
&\Sigma=\frac{\rmi}{2}P^2(\partial\bL-\bpartial L)=
-\frac{P^2}{2}\sigma\end{split}\ee 
$P$, $m$ and $L$ are two real and one complex function
of variables $u,\xi,\bxi$. $\partial$ is defined by \eqref{de}.

The field equations \eqref{eqpr} reduce to \cite{KR}
\be M=P^2\mathrm{Re}\left[\partial\bpartial\Sigma-
2(\partial\Sigma)\bL,_u-\Sigma(\partial\bL),_u+
2\Sigma(\partial\bpartial\log P-\partial\bL,_u)\right]\label{eqalsnut} \ee
and \be \partial(m+\rmi M)=3(m+\rmi M)L,_u.\label{eqals3} \ee
There is an additional equation for vacuum metrics,
\be\label{eqalsvac}\begin{split}
&[P^{-3}(m+\rmi M)]_{,u}=\\
&P[\partial+2(\partial\log P-L_{,u})]\partial
[\bpartial(\bpartial\log P-\bL_{,u})+
(\bpartial\log P-\bL_{,u})^2].\end{split}\ee

We assume $\xi$ to be a complex stereographic coordinate on $S_2$.
Metrics \eqref{als_m}--\eqref{als_c} are asymptotically flat at future
null infinity provided the form $\kappa$, and the functions $m$,
$\hP:=(1+\frec{1}{2}\xi\bxi)^{-1}P$ and $M$ are regular on $R\times
S_2$ \cite{TN}. The Penrose conformal factor is
\be\label{omega}\Omega=\hP r^{-1}.\ee
In \cite{TN} we gave explicit formulae for the modified mass
aspect, Bondi news and angular momentum aspect for metrics given by
\eqref{als_m}--\eqref{als_c}. These quantities correspond to
approximate Bondi coordinates given by
\be\label{BSalsU}\begin{split}
&u'=\hu + \Or (r^{-1}),\\
&r'=\hP^{-1}r-\eta+\Or (r^{-1}),\\ 
&\xi'=\xi + \Or (r^{-1}),\end{split}\ee
where $\hu=\int\hP\rmd u$ and $\eta$ can be found explicitly 
using $\hu$ \cite{TB}. 

It is convenient to introduce the potential $V$ for the
function $P$ \cite{KR,RR}, so that \be\label{potential}P=V,_u.\ee
Equation \eqref{eqalsvac} is equivalent to
\be\begin{split}
&\hP^{-1}[\hP^{-3}(m+\rmi M)-(1+\frec{1}{2}\xi\bxi)^3
\partial^2 \bpartial^2 V],_u=\\
&=-\hP^{-1}(\partial^2 V),_u \hP^{-1}
(\bpartial^2 V),_u=-\hP^{-2}I\bar{I},\end{split}\label{mlossV}\ee
where
\be\label{I}I=P^{-1}(\bpartial^2 V),_u.\ee
For algebraically special solutions \eqref{als_m}--\eqref{als_c}
the potential $V$ \eqref{potential} is
\be V=(1+\frec{1}{2}\xi\bxi)\hu=\int P\,\rmd u,\label{Vrr}\ee
Note that $\hP^{-1}\partial_u$ is an approximation of the derivative
with respect to the Bondi time $u'=\hu+\Or (\Omega)$. One can express
the tensors $n_{AB}$ and $\dot{n}_{AB}$ (see \cite{TB,TN}) in terms of $V$
obtaining
\be\begin{split}
n_{\xi\xi}&=-2(1+\frec{1}{2}\xi\bxi)^{-1}\partial^2 V,\quad
n_{\xi\bxi}=0,\quad n_{\bxi\bxi}=\overline{n_{\xi\xi}},\\
\dot n_{\xi\xi}&=-2P^{-1}\partial_u\partial^2 V=-2\bar{I},\quad
\dot n_{\xi\bxi}=0.\end{split}\label{news}\ee
Now one can interpret the real part of \eqref{mlossV} as the
quasi-local mass loss formula for the twisting algebraically special
metrics and express $\hat M$ in terms of $V$ in the following way:
\be\hat M=m\hP^{-3}-(1+\frec{1}{2}\xi\bxi)^3\mathrm{Re}
(\partial^2\bpartial^2 V).\label{mhatV}
\ee
The imaginary part of \eqref{eqalsvac} defines the NUT parameter
$M$:
\be\label{M=ImddddV} M = 
P^3\mathrm{Im}(\partial^2\bpartial^2 V).\ee

Condition \eqref{amric} is satisfied by all algebraically special
metrics for which \eqref{eqpr} holds. The angular momentum aspect
is given by
\be\label{kappaals}\begin{split}
\kappa_{\xi}=&\frac{1}{16}(\partial -\hat P ^{-1}\partial\hat u
\partial_u)n^2+2(m+\rmi M)\hat P ^{-3}\partial\hat u-\\
&-2(1+\frec{1}{2}\xi\bxi)^4
\left[\partial I(\partial\hat u)^2+\frac{1}{3}\hat P ^{-1}
I_{,u}(\partial\hat u)^3\right].\end{split}\ee

\section{Bondi mass and angular momentum}
It follows from \eqref{kerrschild} that one can obtain the
Kerr--Schild metric as a special case of algebraically special metric
subject to the gauge condition $P=1$ (we identify $v+\mathrm{Re}(
\partial\bL$) with $r$.) Although this gauge is singular from the point
of view of asymptotic flatness, it is convenient to stick to it while
calculating the Bondi mass aspect. In this case the potential $V$ is
just $u$ which implies that the second term in the l.h.s. of
\eqref{mlossV} vanishes by virtue of \eqref{dl=0}:
\be\label{mhdl}-(1+\frec{1}{2}\xi\bxi)^3\partial^2\bpartial^2 V=
(1+\frec{1}{2}\xi\bxi)^3\partial^2\bpartial\bL=0.\ee
From \eqref{M=ImddddV} and \eqref{mhdl} we see that the NUT
parameter $M$ vanishes.
At this point it is worth to note that the latter condition
combined with
Einstein equations restricts the possible congruences \eqref{kong}.
One is left with a complex equation \eqref{eqals3} for real $m$
and \eqref{eqalsvac} (with modified r.h.s. in the non-vacuum case).
Vanishing $M$ enables one to integrate
\eqref{eqals3} using the commutation
relation $[\partial,\bpartial]=-i\sigma\partial_u$ obtaining
\be\label{ms3}m=f\sigma^{-3},\ee
where $f$ is a function independent of $u$ \cite{KR,NTL}.

From \eqref{dl=0} and \eqref{news} we conclude
that both tensors $n_{AB}$ and $\dot{n}_{AB}$ vanish. This implies that the
Bondi news function is zero which is also true in the regular gauge
$\hP=1$ (condition $I=0$ is invariant under all BMS
transformations.) This is a direct proof of absence of the ``true''
gravitational radiation in the Kerr--Schild spacetimes.  The change of
gravitational energy or angular momentum can be caused only by
radiation of matter (e.g. by nonvanishing $\Phi$ in \eqref{eqpr}.)

To use the formulae \eqref{mlossV} and \eqref{kappaals} in
regular gauge $\hP=1$ we introduce new coordinates:
\be\begin{split}
&R= (1+\frec{1}{2}\xi\bxi)r,\\
&U=(1+\frec{1}{2}\xi\bxi)^{-1}u
\end{split}\ee
($\xi$ remains unchanged). This transformation does not change $I$.
Assume that $m$ and $\kappa$ are regular. The Bondi mass is simply
\be\label{mh}\hat{M}=m.\ee
Expression for the angular momentum aspect is also simple
due to $n_{AB}=0$:
\be\label{kappaks}
\hat{\kappa}_{\xi}=\kappa_{\xi}=-2mL.\ee
The field generating null fibres of $\scri$ is given by \cite{TN}
\be\label{vgen}\mathrm{v}=\hP^{-1}\partial_{U},\ee
so the gauge $\hP=1$ allows us to choose $U$ as an approximate
retarded time. 
The approximate Bondi--Sachs coordinates are given by
\be\label{bscks}
\begin{split}
&r'=R-(1+\frec{1}{2}\xi\bxi)^2\mathrm{Re}(\partial\bL)+\Or(r^{-1}),\\
&u'=U+\Or(r^{-1}),\\
&\xi'=\xi +\Or(r^{-1}),\end{split}\ee
where $R=v+\mathrm{Re}\,\partial\bL$.
We can now construct the total energy--momentum
and angular momentum at $\scri$:
\be\label{eam}\begin{split}
&P^{\mu}(u')=\frac{1}{4\pi}\int\limits_{S_2}^{}
ml^{\mu}\rmd\sigma,\\
&L^{\mu\nu}(u')=\frac{3}{8\pi}\int\limits_{S_2}^{}
m(1+\frec{1}{2}\xi\bxi)^2
\Big(l^{[\mu}l^{\nu]},_{\xi}\bL+\cc\Big)\rmd\sigma,\end{split}\ee
where
\be\label{l}
l^{\mu}=(1+\frec{1}{2}\xi\bar\xi)^{-1}
\left[1+\frec{1}{2}\xi\bar\xi,
\sqrt{2}\,\mathrm{Re}\,\xi,
\sqrt{2}\,\mathrm{Im}\,\xi,
1-\frec{1}{2}\xi\bar\xi\right]\ee
and $\rmd\sigma = i(1+\frac{1}{2}\xi\bxi)^{-2}\rmd\xi\iz\rmd\bxi$.

\section{Example: ``radiating Kerr'' metric}
In this section we apply our results to the non-stationary
generalization of the Kerr solution given in the Kerr--Schild form by
Vaidya and Patel in \cite{VP}, and published earlier by Kramer
\cite{KRRK}. Contrary to the observation in \cite{VP}, the
metric is of Petrov type $I\!I$ and not type $D$. As shown in \cite{WIL},
rotating spacetimes of type $D$ with pure radiation do
not exist.

Following \cite{VP} we express the metric in terms of the
Kerr--Schild coordinates defined by
\be\label{Vcart}\begin{split}
x&=[(r+b/2)\cos\phi + a U(u,\theta)\sin\phi]\sin\theta,\\
y&=[(r+b/2)\sin\phi - a U(u,\theta)\cos\phi]\sin\theta,\\
z&=r\cos\theta,\\
t&=u + x\sin\theta\cos\phi + y\sin\theta\sin\phi + z\cos\theta.
\end{split}\ee
The metric takes the following form:
\be\begin{split}\label{Vsol}
\tilde{g} &= \rmd t^2-\rmd x^2-\rmd y^2-\rmd z^2 -
\frac{2mU^{-3}(r+\frec{b}{2}\cos^2\theta)}
{(r+\frec{b}{2}\cos^2\theta)^2+a^2 U^2 \cos^2\theta}\times\\
&\times\left(\rmd t - \frac{z}{r}\rmd z -\frac{(r+\frec{b}{2})
(x\,\rmd x+y\,\rmd y)+ a U(x\,\rmd y - y\,\rmd x)}
{\left(r+\frec{b}{2}\right)^2+a^2 U^2}\right)^2\end{split}\ee
with
\be\label{HiU}
U(u,\theta)=\sqrt{1+\frec{b}{a^2}u -\frec{b^2}{4a^2}\cos^2\theta}.\ee
Here $m$, $a$ and $b$ are constants and
\be\label{kappaVP}
\kappa=\rmd t - \frac{z}{r}\rmd z -\frac{(r+\frec{b}{2})
(x\,\rmd x+y\,\rmd y)+ a U(x\,\rmd y - y\,\rmd x)}
{\left(r+\frec{b}{2}\right)^2+a^2 U^2}
\ee
is the form corresponding to the null geodesic congruence. Note that
the metric \eqref{Vsol} rewritten in terms of coordinates
$(u,r,\theta,\phi)$ is axisymmetric and in the limit $b\to 0$ one
obtains the Kerr metric in the Kerr--Schild form.

Observe that $\Omega = r^{-1}$ can be chosen as the Penrose conformal
factor. This means that $\hP=1$, so we choose $u$ to be an approximate
Bondi time. For $u>b/4-a^2/b$ the function $U$ is well defined for all
$\theta\in[0,\pi[$. Regularity of the unphysical metric $\Omega^2\tilde{g}$ in
the neighbourhood of $\scri$ is obvious. From these observations we
conclude that the ``radiating Kerr'' metric is asymptotically flat at
a piece of $\scri$.

The function $U$ satifies the equations
\be\begin{split}\label{veqM}&\left[a^2+bu-(b/2a)^2\cos^2\theta\right]
\partial_u (mU^{-3})=-\frac{3}{2}b(mU^{-3}),\\
&\frac{b}{2}\sin\theta\cos\theta\,\partial_u (mU^{-3})=
\partial_{\theta}(mU^{-3})\end{split}\ee (equations (3.4) and (3.3) in
\cite{VP}.) The first is the mass loss formula \eqref{mloss} with
$\dot{n}_{AB}$ put equal to zero. The second is equivalent to \eqref{eqals3}
expressed in terms of $u$, $\theta$ and $\phi$, with vanishing NUT
parameter.

Rewriting \eqref{kappaVP} in terms of $u$,
$\theta$ and $\phi$ we get \be\label{kappa_utp}
\kappa=\rmd u + \frac{b}{2}\sin\theta\cos\theta\,\rmd\theta-
aU\sin^2\theta\,\rmd\phi.\ee
Next, calculating
\be\label{twist}\kappa\iz\rmd\kappa=-2aU\cos\theta\,\rmd u\iz\mathrm{vol}_{S_2},
\ee
we find from \eqref{als_c} that $\Sigma=aU\cos\theta$. This
shows, in agreement with \eqref{veqM} and \eqref{ms3} that 
$m U^{-3}$ plays a role of $m$ in \eqref{als_c}.

\subsection{Modified Bondi mass aspect and total energy}
Using formula \eqref{mh} we get the following expression for the
Bondi mass:
\be\label{mhatVa} M(u,\theta)=\frac{m}{U^3}=
\frac{m}{\left[1+bu/a^2-(b\cos\theta/2a)^2\right]^{3/2}}.\ee
Averaging both sides of \eqref{mhatVa} over $S_2$
gives the total energy:
\be\label{EV}E(u) = \frac{m}{[1+bu/a^2-(b/2a)^2]^{1/2}(1+bu/a^2)}.\ee
It is obvious that $E$ is a decreasing function of $u$ and that
$\lim_{u\to\infty}E(u)=0$ in agreement with the fact that in this
limit the metric becomes flat. It is easy to check using
\eqref{eam} and \eqref{mhatVa} that the spatial part of the
energy--momentum four--vector vanishes. It follows from \eqref{HiU}
and \eqref{EV} that $E(u)$ is well defined if $U$ is well defined.

\subsection{Angular momentum}
The spherical coordinates $\theta$ and $\phi$ define the
stereographic coordinate used in \eqref{als_m}:
\be\label{stereo}
\xi=\sqrt{2}\tan(\theta/2)\exp(\rmi\phi).\ee
The function $L$ can be found as a coefficient
in the decomposition
\be\label{kappacV}
\kappa-\rmd u=\frac{b}{2}\sin\theta\cos\theta\rmd\theta-
aU\sin^2\theta\rmd\phi=L\,\rmd\xi+\bL\,\rmd\bxi .\ee
The result is
\be\label{LV} L=\frac{\bxi}{(1+\frec{1}{2}\xi\bxi)^2}
\left(\rmi a U + \frac{b}{2}\frac{1-\frec{1}{2}\xi\bxi}{1+\frec{1}{2}\xi\bxi}
\right).\ee
Now the formula \eqref{kappaks} applied to ``radiating Kerr'' metric
gives, after returning to the spherical coordinates,
the following expression for angular momentum aspect:
 \be\label{kappaV}\kappa_A\rmd x^A =
\frac{2ma\sin^2\theta}{U^2}\rmd\phi
-\frac{mb \sin\theta\cos\theta}{U^3}\rmd\theta.\ee
Denoting $(\theta,\phi)$ by $(x^2,x^3)$ we obtain
the ``tensorial'' part of the integrand in (\ref{angm}) in the following
form:
\be\label{lmlna}\begin{split}
l^{[\alpha}_{\phantom{A}}
l^{\beta]}_{\ \,\,|A}&=-\frac{1}{2}\sin\theta\,\delta^{[\alpha}_0
\delta^{\beta]}_3\,\delta^2_A+\frac{1}{2}\sin^2\theta\,
\delta^{[\alpha}_1\delta^{\beta]}_2\,\delta^3_A+\\
&+\textrm{terms proportional to}
\,\,\cos\phi\,\,\textrm{or}\,\,\sin\phi.\end{split}\ee
Integrating the terms containing $\sin\phi$ or $\cos\phi$
over $S_2$ with some axisymmetric function gives zero.
The same holds for the first term in \eqref{lmlna}.
Inserting \eqref{kappaV} and \eqref{lmlna} into \eqref{angm} yields
the following form of the angular momentum tensor:
\be\label{amV} L^{\mu\nu}=ma\,\delta^{[\mu}_1\delta^{\nu]}_2
\frac{3a^2}{b^3}\left[
2b - \frac{4a^2-b^2+4bu}{\sqrt{a^2+bu}}\mathrm{atanh}
\frac{b}{2\sqrt{a^2+bu}}\right].\ee
The only nonvanishing component is $L^{12}$.
Formula \eqref{amV} can be expanded in the powers of $b$ around
$b=0$:
\be\label{LVexpb} L^{\mu\nu}=L^{\mu\nu}_{\mathrm{Kerr}}
(1-bu/a^2)+\Or (b^2).\ee
Note that, as expected, $\lim_{b\to 0}L^{\mu\nu} = L^{\mu\nu}_{\mathrm{Kerr}}$
and $\lim_{a\to 0}L^{\mu\nu} = 0.$
Similarly to $E(u)$, $L^{\mu\nu}(u)$ is well defined if $U$ is well defined.
Differentiating $\kappa_{\phi}$ from \eqref{kappaV}
with respect to $u$ we obtain some negative function,
hence $L^{\mu\nu}$ is a decreasing function of $u$. As in the
case of total energy, all angular momentum is radiated away
by pure radiation so that $\lim_{u\to\infty}L^{\mu\nu}(u)=0$.

\section*{Conclusions}
In this paper we discussed the Kerr--Schild metrics from the point of
view of asymptotic flatness at future null infinity. We found the
approximate Bondi--Sachs coordinates as well as formulae expressing
total energy--mome\-ntum and angular momentum at $\scri$.
Bondi mass and angular momentum aspect are simple functions of the
metric components. Our ``negative''
result shows that the Kerr--Schild metrics cannot be used to construct
realistic models of gravitationally radiating systems.
We gave an example of calculation of the total energy and angular
momentum for exact Kerr--Schild solution presented by Vaidya and
Patel. Our results cannot be directly compared with those obtained by
Kramer and von der G{\"o}nna \cite{KG} because of other sections
of $\scri$ on which the mass and angular momentum integrals have
calculated. The difference is caused by other definition of retarded time.
The transformation between retarded time used here and that used in
\cite{KG} is a supertranslation. 

\section*{Acknowledgments}
I am grateful to Professor J. Tafel for helpful discussions
and critical remarks.
I highly appreciate lectures of Professor A. Trautman on
differential geometry and spinors.


\begin{thebibliography}{25}
\bibitem{TR} Trautman A 1962
{\sl On the propagation of information by waves}, in: Recent
Developments in General Relativity, Pergamon Press---PWN, p. 459

\bibitem{KS} Kerr R P, Schild A 1965
%{\sl Some algebraically degenerate solutions of Einstein
%gravitational field equations,}
{\it Proc. Symp. Appl. Math.} {\bf 17} 199

\bibitem{GG} G\"urses M, G\"ursey F 1975
%{\sl Lorentz covariant treatment of the Kerr--Schild geometry,}
{\it J. Math. Phys.} {\bf 18} 2356

\bibitem{CHS} Coll B, Hildebrandt S R, Senovilla J M M 2001
%{\sl Kerr--Schild Symmetries}
{\it Gen. Rel. Grav.} {\bf 33} 649

\bibitem{HI} Hildebrandt S R 2002
%{\sl A Physical Application of Kerr--Schild Groups}
 {\it Gen. Rel. Grav.} {\bf 34} 65 

\bibitem{GS} G\"urses M, Sar{\i}o{\u g}lu \"O 2004
%{\sl Some Further Properties of the Accelerated Kerr--Schild Metrics}
{\it Gen. Rel. Grav.} {\bf 36} 403

\bibitem{CA} Cadavid A C, Finkelstein R J 1999
%{\sl Kerr--Schild Description of a Rotating Dyon}
{\it Gen. Rel. Grav.} {\bf 31} 31

\bibitem{TB} Tafel J 2000
%{\sl Bondi mass in terms of the Penrose conformal factor,}
{\it Class. Quantum Grav.} {\bf 17} 4397

\bibitem{CM} Cornish F H J, Micklewright B 1996
%{\sl Photon rockets and gravitational radiation,} 
{\it Class. Quantum Grav.} {\bf 13} 2499

\bibitem{KIN} Kinnersley W 1969
%{\sl Field of an arbitrarily accelerating point mass,}
{\it Phys. Rev.} {\bf 186} 1335

\bibitem{KGpr} Kramer D, von der G\"onna U 1998
%{\sl Rotating pure radiation,}
{\it Class. Quantum Grav.} {\bf 15} 215

\bibitem{VP} Vaidya P C, Patel L K 1973
%{\sl Radiating Kerr Metric,}
{\it Phys. Rev.} D {\bf 7} 3590

\bibitem{KG} Kramer D, von der G\"onna U 1998
%{\sl Rotating pure radiation field of the Kerr--Schild type,}
{\it Class. Quantum Grav.} {\bf 15} 2017

\bibitem{KRRK} Kramer D 1972
%{\sl Gravitational field of a rotating radiating source},
3rd Sov. Grav. Conf. Erevan, Tezisy, p. 321.

\bibitem{TN} Natorf W, Tafel J 2004
%{\sl Asymptotic flatness and algebraically special metrics,}
{\it Class. Quantum Grav.} {\bf 21} 5397

\bibitem{TW} Tamburino L A, Winicour J H 1966
%{\sl Gravitational fields in finite and conformal Bondi frames,}
{\it Phys. Rev.} {\bf 150} 1039

\bibitem{JEZ} Jezierski J 1998
%{\sl Trautman--Bondi mass in classical field theory,}
{\it Acta Phys. Polon.} B {\bf 29} 667

\bibitem{DS} Dray T, Streubel M 1984
{\it Class. Quantum Grav.} {\bf 1} 15

\bibitem{KR} Stephani H, Kramer D, MacCallum M A H,
Hoenselaers C and Herlt E 2003
{\it Exact Solutions to Einstein's Field Equations, Second Edition,}
(Cambridge: Cambridge University Press)

\bibitem{RR} Robinson I, Robinson J R 1969
%{\sl Vacuum metrics without symmetry,}
{\it Int. J. Theor. Phys.} {\bf 7} 231

\bibitem{NTL} Lewandowski J, Nurowski P and Tafel J 1991
%{\sl Algebraically special solutions of the Einstein
%equations with pure radiation fields,}
{\it Class. Quantum Grav.} {\bf 8} 453

\bibitem{WIL} Wils P 1990
%{\sl Aligned pure radiation of type D does not exist,}
{\it Class. Quantum Grav.} {\bf 7} 1905 (comment)

\end{thebibliography}
\end{document}